\documentclass[epj]{svjour}
\usepackage{graphics}
\begin{document}

\title{Enhancement of annihilation cross sections by electric 
interactions between 
the antineutron and the field of a large nucleus. 
} 

\author{A.~Bianconi\inst{ }, E. Lodi Rizzini\inst{ }, 
V.Mascagna\inst{ }, and L.Venturelli\inst{ }}
\institute{Dipartimento di Ingegneria dell'Informazione, 
Universit\`a degli Studi di Brescia, I-25133 Brescia, Italy, 
\\and 
Istituto Nazionale di Fisica Nucleare, Gruppo Collegato di Brescia 
\\Email: andrea.bianconi@unibs.it}

\abstract{
Data relative to antineutron and antiproton annihilation on large 
nuclei in the range 75-200 MeV/c present two unexpected 
features: (a) antineutron  
and antiproton cross sections have a similar size, (ii) the rise of 
the antineutron cross section at decreasing energy is much steeper than 
predictable for an inelastic process of purely strong nature at 
that energy. The observed behavior of $\bar{n}$-nucleus annihilations 
is similar to what would be expected for $\bar{p}$-nucleus annihilations, 
where Coulomb attraction focusses $\bar{p}$ 
trajectories towards the nucleus, enhancing the inelastic cross section 
by a factor $O(1/p)$ with respect to $\bar{n}$ on the 
same target. This results in a $1/p^2$ behavior at small energies. 
The presence of a similar enhancement in the antineutron 
case may only be justified by an interaction with a longer range than 
strong interactions. Excluding a Coulomb force because of the $\bar{n}$ 
neutrality, and taking into account that an intrinsic electric dipole 
is forbidden for the antineutron, the next choice is an electric dipole 
that is induced by the nuclear electric field. Recent theoretical works 
have shown that a non-negligible electric polarization may be 
induced in a neutron by QED vacuum polarization. Assuming this as 
a possibility, we have used a simple model to calculate the polarization 
strengths that are needed to fit the available data in terms of this 
effect. These are within the magnitude predicted by the vacuum 
polarization model. We have also discussed alternative scenarios that 
could induce an electric polarization of the antineutron as a consequence 
of the interplay between strong and e.m. interactions. 
} 

\PACS{
 {12.20.-m}{Quantum electrodynamics}\and
 {24.80.+y} {Nuclear tests of fundamental interactions and symmetries}\and
 {25.43.+t} {Antiproton-induced reactions}
}


\maketitle

\section{Introduction}
\label{intro}

\subsection{Induced electric dipole in an antineutron} 
\label{induced}

Although the neutron (and equivalently the antineutron) is  
composed by constituents with nonzero electric charge, the existence 
of an intrinsic electric dipole moment for this particle 
is forbidden by general arguments. 
If the neutron had an electric dipole moment, P or T 
transformations would exchange the relative orientation of its 
magnetic and electric moments. 

Since this rule is only valid in a P- or CP-conserving context,  
many experiments have been performed 
to establish upper limits to this 
feature (see \cite{Baker06,Fedorov10} for recent measurements, and 
\cite{Dar00,Pospelov05,Peng08,Lamoreaux09} for reviews of 
the theoretical background 
and of the previous measurements). If an intrinsic dipole exists, 
it is very small. 

In a CP-conserving context, a non-elementary particle may 
be polarized by an external electric field. No strict rule forbids 
this, because the relative orientation of the magnetic 
and electric moments in 
this case is not an intrinsic permanent property of the 
hadron. The most intuitive mechanism producing a polarization, that is 
a constituent charge displacement, requires mixing neutron internal 
states with opposite parity (see Sect.~\ref{discussion} for details). 
This may take place if these states are almost degenerate in mass, while  
the first negative parity state is several hundred MeV over the 
fundamental level. 

In a recent theoretical work \cite{Dominguez09} 
it has been shown that QED corrections of order $\alpha^2$ (in the 
amplitude) permit an indirect interaction between the field of the 
neutron magnetic moment and an external electric 
field, mediated by vacuum polarization. As a consequence, the neutron 
behaves as if an electric dipole with strength proportional to the 
inducing field strength were present. This does not require 
displacements in the internal neutron charge distribution (this would be 
a lowest order QED effect). So it is a possible effect for 
an antineutron in its ground state. However, it requires an 
electric field whose strength is near the magnitude 
10$^{18}$ V/m (Sauter's critical field \cite{Sauter31}). 
The calculation of \cite{Dominguez09} is aimed at laser fields and 
motivated by the hope that lasers may approach this value in a near 
future. 
A later work \cite{Zimmer12} applies the same ideas to neutron 
elastic scattering by nuclei. 

At the surface of a heavy nucleus the electric field has magnitude 
10$^{21}$ V/m and a correspondingly large gradient. If 
an electric dipole may be induced in a neutron or an antineutron, 
short distance projectile-nucleus collisions 
are the place where this mechanism is strongest in our present 
possibilities. 
It is not clear to us at which extent the predictions 
by \cite{Dominguez09} and \cite{Zimmer12} may be adapted to the 
inelastic nuclear physics processes 
considered in the following. However, these give us a chance to 
consider the $possibility$ that a response of some kind 
is induced in a slow antineutron by the electric field of a heavy nucleus.  
This would enhance the antineutron annihilation cross section at 
small energies, 
much as the Coulomb force does with antiprotons. 

\subsection{Antineutron-nucleus and antiproton-nucleus 
annihilation cross section at small energies} 
\label{antineutron1}

The data that are most relevant for the present work are the cross sections 
of antineutron annihilation on medium and large nuclei 
for projectile momentum $p$ $<$ 400 MeV/c. These have been measured by 
several collaborations in the years 1980-2002 
\cite{Gunderson81,Agnello88,Barbina97,Ableev94,Astrua02} 
(see also \cite{Bressani03} for a state-of-the-art review). The  
$\bar{n}$-nucleus annihilation data of \cite{Astrua02} 
(see table 2 and fig.7 in that work) cover the range 76-380 MeV/c 
on several nuclear targets 
from carbon to lead, and are presently the available data reaching 
the smallest energy. 
For $p$ $\geq$ 128 MeV/c, the data from this 
experiment confirm very well previous data  
\cite{Ableev94} and are almost perfectly fitted by an  
$a+b/p$ law, with a relevant contribution of the $b/p$ term. 
The smallest energy point at 76 MeV/c overrates by 15-20 \% 
this fit, for all the target nuclei, and an additional 
$1/p^2$ term is needed to fit this last point. 

In the case of the antiproton annihilations on light and heavy nuclei 
at low energies, many measurements have been performed in the region of 
momentum $p$ $<$ 400 MeV/c (see 
Ref. \cite{bizzarri1974,kalogeropoulos1980,balestra1984,balestra1985,balestra1986,balestra1989,bruckner1990,bertin1996,benedettini1997,zenoni1999_1,zenoni1999_2,bianconi2000_1,bianconi2000_2,Bianconi11}). 


For $p$ $>$ 200 MeV/c antineutron and antiproton cross sections 
on the same target are almost equal (see fig.11 in \cite{Astrua02}). 
When momenta decrease below 200 MeV/c one expects $\bar{p}$-nucleus 
annihilation cross sections to grow much larger than antineutron ones, 
because of the Coulomb field of the nucleus pulling antiprotons towards 
the annihilation region from long 
distances \cite{LL3,Bianconi00,Bianconi00b,Batty01,Carbonell93}.  

This dominance 
is well verified with data on light nuclear targets (in particular, 
with hydrogen targets, see fig.1 in \cite{Friedman14}), but has not 
been seen yet in  annihilations on medium/large nuclear targets down 
to 76 MeV/c. 
On the contrary, a compared analysis 
of $\bar{p}$-Sn 
\cite{Astrua02} and $\bar{n}$-Sn \cite{Bianconi11} annihilation 
data below 100 MeV/c, suggests that a low energy antineutron behaves 
in a more ``electrostatic'' way than an antiproton (see fig.6 in 
\cite{Friedman14}). 

Even ignoring the comparison between $\bar{p}$ and $\bar{n}$ data, 
it is difficult to explain the steep rise of the antineutron 
cross sections in the region 76-200 MeV/c in absence 
of medium/long range interactions, that is of interactions whose 
range is larger than the standard range of strong interactions. 
This point is better examined in Sect.~\ref{discussion}. 


In the following we assume that this interaction is of electromagnetic 
origin. Excluding Coulomb $\bar{n}$-nucleus interactions 
because of the overall neutrality 
of the antineutron, several possibilities are left, none of them 
obvious. The simplest one is an induced electric dipole. 
In Sect.~\ref{discussion} we will analyse its possible origin from several points 
of view. For the time being we just assume that it exists and use it 
to improve a basic fit of the data. 



\section{Calculation method} 
\label{calculation}

In the case of Coulomb interactions, the cross section enhancement with 
respect to a black sphere model may be calculated in a semiclassical way 
(see eq. 22 of \cite{Batty01}). With an electric dipole force the 
mathematics is more complex, so we will work numerically starting 
from a classic trajectory bending model. The procedure is the following. 

1) The centre of the target nucleus is in the origin, and the projectile 
antineutron moves from the position $(0,\ y_o,\ -\infty)$ with 
momentum $(0,0,p)$. 

2) We assume that in absence of dipole effects 
the annihilation only involves antineutrons 
whose momentum is parallel to the $z$ axis until they reach the 
annihilation region. This region is assumed to be a sphere with 
radius $R_{int} = 1.2 \cdot A^{1/3}$ fm. That is to say, only those antineutrons 
that ``touch'' the target nucleus may be involved in the annihilation 
process. This regards antineutrons with initial impact parameter 
$y_o$ $\leq$ $R_{int}$. 

3) We assume that the interaction between these antineutrons and the 
nucleus is responsible for a cross section of the form 
$a$ $+$ $b/p$ exactly as given in eq.3 of \cite{Astrua02}, 
with the values of 
the parameters $a$ and $b$ from table 3 of that 
work ($a = 66.5 \pm 3.0$ mb, $b = (1.987 \pm 0.086) \cdot 10^4$ mb MeV/c). With all the used target nuclei this fit reproduces very well 
the measured cross section with the exception of the lowest point 
at 76 MeV/c. 

4) We introduce a long-distance force of the form $F(r)$ $=$ 
$D(r) \partial E(r)/\partial r $, 
where $E(r)$ $=$ $Ze/(4\pi\epsilon_o r^2)$ is the nuclear field 
and $D(r)$ $\equiv$ $\beta E(r)$ is the induced dipole. The polarizability 
$\beta$ is a constant parameter to be fitted. 

5) We calculate 
the classical trajectories of the antineutrons in this field, out 
of the strong interaction region. 
since this field focusses projectile trajectories towards the target, 
the antineutrons reaching a 
target sphere of radius 1.2 $A^{1/3}$ fm have initial impact parameter 
$y_o$ $\leq$ $R_{dip}$, with $R_{dip}$ $>$ $R_{int}$. 

6) The fit $a + b/p$ (cross section without dipole) 
is multiplied by the correcting factor 
$R_{dip}^2 / R_{int}^2$. This gives the ``dipole-corrected cross section''. 
We assume that this is the correct value of the 
cross section below 100 MeV/c. 

7) This procedure is repeated for several values of the polarization 
constant $\beta$, until we find a satisfactory reproduction of all the 
data of \cite{Astrua02} for one nuclear target. 

8) This calculation has been performed for the cases of Cu and Sn targets 
($Z$ $=$ 29 and 50 respectively, $A$ $\approx$ 64 and 119 respectively 
after averaging over isotopes).  
Considering the very regular dependence of the data of \cite{Astrua02} on 
the nuclear mass, we have no reasons to imagine that repeating the 
calculation for more targets would present surprises.

\section{Results} 
\label{results}

\begin{figure}[ht]
\resizebox{0.50\textwidth}{!}{%
\includegraphics{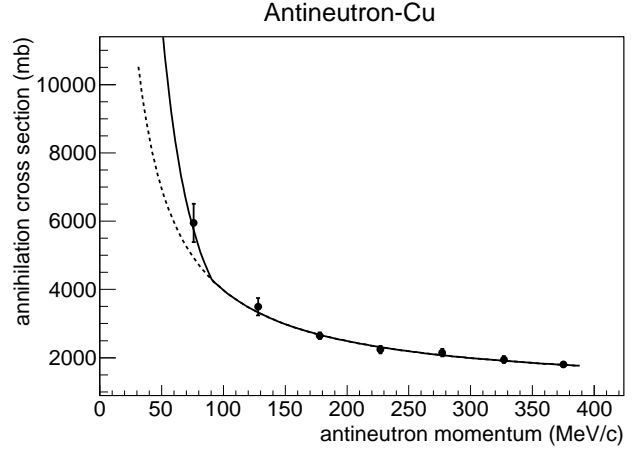}}
\caption{
Points with error bars: $\bar{n}$Cu annihilation data from \cite{Astrua02}. 
Dotted line: fit of the form $a + b/p$, with $b/a$ from table 3 of 
\cite{Astrua02}. Continuous line: dipole-corrected fit (see text). 
Remark: where the dotted line is not visible (for $p$ $>$ 100 MeV/c), 
it is because of perfect overlap with the other curve. 
\label{fig:f1}
}
\end{figure}

\begin{figure}[ht]
\resizebox{0.50\textwidth}{!}{%
\includegraphics{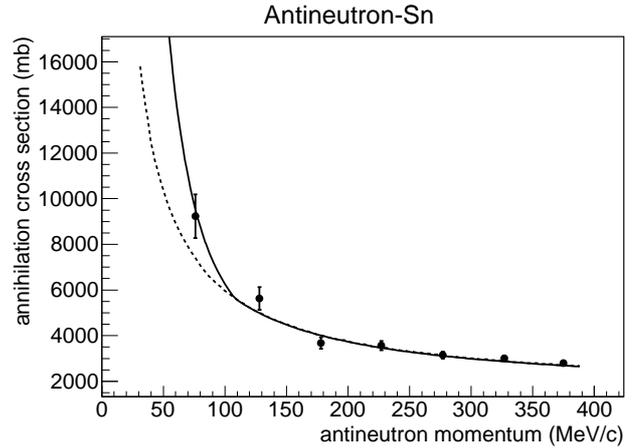}}
\caption{
Points with error bars: $\bar{n}$Sn annihilation data from \cite{Astrua02}. 
Dotted line: fit of the form $a + b/p$, with $b/a$ from table 3 of 
\cite{Astrua02}. Continuous line: dipole-corrected fit (see text). 
Remark: where the dotted line is not visible (for $p$ $>$ 100 MeV/c), 
it is because of perfect overlap with the other curve. 
\label{fig:f2}
}
\end{figure}

The results are presented in figures \ref{fig:f1} ($\bar{n}$-Cu) 
and \ref{fig:f2} ($\bar{n}$-Sn). In each figure we report:   

a) The data points by \cite{Astrua02}, and 
their $a+b/p$ fit that is very good on all the points 
for $p$ $>$ 100 MeV/c.  

b) Our dipole-corrected fit that allows fitting the 75 MeV/c point without 
including an additional $1/p^2$ term. 

To give an understandable meaning to the fitted polarizability, 
we redefine  
the induced dipole moment as 
\begin{equation}
D(r)\ = \beta E(r)\ 
\equiv\ e Z L\ {{1 fm^2} \over r^2}.
\end{equation}
where 
$r$ is the distance of the antineutron to the nucleus center. 
If 
the nuclear electric field were emitted by a pointlike particle with 
charge $Ze$, at a distance 1 fm from this particle the induced dipole would 
assume the value $eZL$, physically readable as an electron and a positron    
separated by a distance $LZ$. 
The fitted values of $L$ are summarized in table 1. 
The uncertainties are due to the error bars in the point at 76 MeV/c 
of each data set, visible in figs. 1 and 2. 

\vspace{0.5truecm}
\begin{table}
\caption{Fitted $L$ parameters. See text for details.}
\label{tab:1}  
\begin{center}
\begin{tabular}
{|l|l|}
\hline
\multicolumn{2}{|c|}{$L$ parameters} \\
\hline
Cu & 3.8 $\pm$ 0.7 fm \\ 
Sn & 1.7 $\pm$ 0.4 fm \\ 
\hline
\end{tabular} 
\end{center}
\end{table}
\vspace{0.5truecm}

At the nuclear surface, where the dipole force is most 
relevant, the induced dipole is $eLZ/R_{nucleus,fm}^2$. 
These values imply 
$D$ $=$ $e\cdot$4.8 fm 
at the surface of a Cu nucleus, and $D$ $=$ 
$e\cdot$2.4 fm 
at the surface of Sn. When these $D$ values are 
multiplied by $\partial E(r) / \partial r$ we find that the corresponding 
forces are roughly the same in the two cases. 

These forces are medium range  
ones, that is longer than strong interactions but shorter than 
Coulomb ones. 
Within 1-2 fm of the nuclear surface they are almost as relevant 
as a Coulomb force: 
compared to the Coulomb force applied to an antiproton, at the Sn surface the 
antineutron 
dipole force is 9/10, about 1/3 at 2 fm off the nuclear surface and 
1/10 at 6 fm. An analysis of the antineutron 
trajectories in our model shows that they are bended by dipole forces 
at distances of several fm, but serious deflections are confined to the 
last 2-3 fm of path before contact with the annihilation region. 

\section{Discussion} 
\label{discussion}

\subsection{Short and medium range interactions in the annihilation 
process} 
\label{short}

The electric dipole force as fitted in our work behaves as 
a ``medium range'' one: if included in an optical model potential 
$U(r)$ $+$ $i W(r)$, it would 
imply a real negative (that is, attracting) term $U(r)$ with range 
about 2 fm larger than the 
imaginary (annihilating) term $W(r)$. 
Forgetting to interpret this interaction as an electric dipole 
force, our fits suggest the following consideration:

the 
$1/p^2$-dependence of the data by \cite{Astrua02} at small momenta, 
and the fact that the absolute value of the cross sections competes  
in magnitude with the corresponding antiproton cross sections, may be 
explained in terms of an elastic and attractive interaction that 
(i) in a 
range extending 2 fm off the nuclear surface has the same size as a 
Coulomb interaction between charges $-e$ and $+Ze$, (ii) becomes 
rapidly negligible at larger distances. 


Is a medium range interaction term really necessary to interpret 
the antineutron data? In our opinion it is, for the following reasons. 
Let us first assume a nonresonant behavior. 
The key parameters to qualitatively discuss this question are:  

1) The diameter $d_{A,strong}$ of the strong interaction region, that is 
the nuclear diameter plus 1 fm because of the range of the strong 
interactions. For Sn this means 12-13 fm, and 
defines a maximum cross section 40 fm$^2$ for the region where the 
interactions directly responsible for the annihilation are effective. 

2) $\hbar/p$, that at 100 MeV/c is 2 fm. This wavelength defines the 
borderline between a semiclassical regime and a fully quantum regime. 
Since the size of the region that is involved in the interaction 
cannot be smaller than the projectile wavelength because of the 
indetermination principle, if $\hbar/p$ $>$ $d_{A,strong}$ we are in 
a fully quantum regime. In this case a nonresonant 
inelastic cross section should 
follow a Bethe $1/p$ law \cite{LL3}.  In 
the opposite regime $\hbar/p$ $<$ $d_{A,strong}$ the cross section 
of the annihilation region cannot exceed much the geometrical cross 
section of the target nucleus. For Sn the borderline is  
$p$ $\approx$ 30 MeV/c, and for $^{12}$C $p$ $\approx$ 70 MeV/c. 
Since these borderlines are not sharp, it 
is reasonable to expect a $small$ $1/p$ contribution at $p$ $\approx$ 100 
MeV/c, but not a steep rise. 

3) The absorbing power of the elementary 
$\bar{n}$-nucleon interaction. 
In the considered energy range, the experimental $\bar{n}$-nucleon 
annihilation 
cross section is 
always $>$ 10 mb. The corresponding mean free path 
$\rho \sigma$ is $<$ 1-2 fm. 
This means that the incoming antineutron flux is exponentially 
damped inside the target nucleus, with a damping range $<$ 1-2 fm. 
With this premise a medium or large nucleus 
is a completely opaque sphere for an $\bar{n}$ flux at 200 MeV/c. 
The systematic increase of the $\bar{n}-p$ annihilation cross 
section below 200 MeV/c \cite{Iazzi00} cannot further increase this (already 
complete) opacity. 


Any model that incorporates the above features will predict an 
almost constant 
$\bar{n}$-nucleus annihilation cross section for 
$\bar{n}$ momenta over 50 MeV/c, of the form $\sigma$ $\approx$ 
$\sigma_{geometrical} + b/p$ with a relatively small $b/p$ term. This 
is also what is is observed, but only for momenta over 200 MeV/c. 

With a $\bar{p}$ projectile, the only difference 
is given by the Coulomb attraction. Since 
$Ze^2/r$ $\approx$ 12 MeV at the Sn surface, and 3 MeV at the C 
surface, relevant effects of this interaction are not expected for 
projectile momenta $>>$ 200 MeV/c. 

If we consider as an example ref.\cite{Friedman14} (see fig.6 of that work) 
and ref.\cite{Bianconi00} (see fig.3 of that work), two independent model 
calculations confirm the previously described picture 
for antiproton and antineutron annihilations on medium-large nuclei 
in the region 50-400 MeV/c. 
Neither may explain the steep observed rise of the antineutron data 
in the momentum range 76-227 MeV/c. 
Is it difficult to imagine a model predicting cross sections that 
$qualitatively$ 
differ from the ones by \cite{Bianconi00} and 
\cite{Friedman14}, as far as this model implements nuclear 
interactions that are short-ranged, present 
homogeneous features on the nuclear volume, and damp the projectile 
wavefunction within a range of 1-2 fm, 
on large nuclei at $p$ $>$ 50 MeV/c. 
So one is obliged to introduce properties beyond this 
baseline modelling. 

These cannot be related to features of one specific target. 
The data of \cite{Astrua02} change in a regular way while passing 
from $^{12}$C to very heavy nuclei.  
In all the considered momentum range including 76 MeV/c, 
the measured cross sections may be fitted as $f(E) \cdot A^x$ 
with $x$ different from 2/3 by less than 3 \% and $f(E)$ that 
does not depend on the target (see fig.9 in 
\cite{Astrua02}).  
Although the neutron/proton ratio of the target nuclei 
ranges from 1:1 (C) to 1.5:1 (Pb), no dependence of the cross section 
on the neutron excess is seen. 
All this suggests that no  effect is present that may be related 
to the features of a few specific nuclei, like a neutron halo. 
A resonance associated with the formation of a coherent 
$\bar{n}-A$ compound nucleus, or a set of overlapping 
resonances, would modify the cross section, but in a  
target-dependent way. 

Resonance peaks could be present in 
the elementary $\bar{n}p$ and $\bar{n}n$ cross sections  
below 50 MeV/c. The corresponding resonance in 
$\bar{n}$-nucleus would present a width 
$\sqrt{\Gamma_{res}^2 + \Gamma_{Fermi}^2}$ 
$>$ 200 MeV/c ($\Gamma_{res}$ is the resonance width, and 
$\Gamma_{Fermi}$ $\approx$ 200 MeV/c is 
the effective width of the nuclear Fermi motion). This means that 
at 200 MeV/c the cross section should already present half of its 
resonance peak value (after subtracting from both the nonresonant 
$a+b/p$ background as fitted at larger momenta). So, the present 
data depend on energy more than a resonance can. 

Summarizing the previous arguments, an explanation of the available 
data needs an interaction whose range is longer that the standard 
Yukawa strong interaction range, and whose strength 
is $\approx$ $e^2/R_{nucleus}$ near the nuclear surface. 
This interaction must be systematically present 
in all the medium/large stable nuclear species, 
depending in a regular way on their gross features.  

So, it is natural to think at electrostatic interactions.  
The point is whether physical mechanisms may 
be imagined, able to produce interactions of this class. 

\subsection{Antineutron polarization via displacement of internal charges}
\label{antineutron2}

We first consider an electric dipole moment induced in the most 
obvious way: an opposite space displacement of the positive and negative 
antiquarks 
composing the antineutron, caused by the electric field surrounding 
the nucleus. 
If the internal position of an antiquark   
is given by the wavefunction $\Psi_L(\vec r)$ with orbital 
angular momentum $L$, no left/right asymmetry is  
present in $|\Psi_L(\vec r)|^2$. Such an asymmetry is present in 
$|a\ \Psi_{L1}(\vec r) + b\ \Psi_{L2}|^2$ if $L1$ and $L2$ are angular 
momenta of opposite parity. Mixing the antineutron ground state 
with an opposite parity excited state requires an excitation energy 
of several hundred MeV. This is impossible for a free or weakly 
interacting antineutron.  

The values of annihilation cross sections reported in 
the quoted references mean a mean path of magnitude less than 1 fm 
before annihilation, for an antineutron inside the target nucleus. 
So, the antineutron is destroyed as soon as it touches 
the nucleus. On the other side, its 
constituent antiquarks keep travelling through the nucleus, binding 
themselves with nuclear quarks and forming mesons that further 
interact with 
the nuclear matter forming more fragments. 

In other words, 
inside the nucleus the wavefunction describing the state of the 3 
antiquarks loses its features of a bound state. 
This removes any constraint on the existence of an electric 
polarization for this set of 3 antiquarks. However, 
inside the nucleus the effects of an electric polarization cannot be 
disentangled from those of the strong interactions. 


The question becomes whether a polarization mechanism that is allowed 
inside the nucleus may be effective at some distance off 
the nuclear surface, where strong interactions are not present. 
A regular interpolation must be present between the two wavefunctions 
describing the antiquarks out of the nucleus (where they form an 
antineutron, with all the described constraints forbidding a low-order 
QED polarization), and the antiquarks inside the nucleus (where no 
such constraints exist). The chances to find an electric polarization 
of an antineutron at some distance from the nuclear surface depend 
on the size of this interpolation region. 

Summarizing this part, an induced electric dipole at lowest order QED 
in an antineutron near the  surface of the target nucleus is not strictly 
forbidden, but a detailed annihilation model 
is required to quantify the strength of this effect. 



\subsection{Antineutron polarization via vacuum polarization} 
\label{antineutron3}

Let us consider the scheme proposed in 
\cite{Dominguez09}. At qualitative level 
this physical picture imagined 
for neutrons in quasi-static laser fields seems 
suitable for the case in study. 

The intense value of the electric field at the Sn surface implies that 
the difference in electrostatic energy $U(r_1)$ $-$ $U(r_2)$ 
is sufficient to create from the 
vacuum a $e^+e^-$ pair within a distance $r_2-r_1$ $\approx$ 
0.5 fm. This vacuum polarization effect is taken into account 
in ordinary charge renormalization in QED, that however assumes 
a pointlike and far test charge. 
Because of its proximity with the source of the nuclear field and of 
its finite diameter $>$ 0.5 fm,  
an antineutron is able to test the local effects of 
this vacuum polarization. 

In the scheme proposed by \cite{Dominguez09} the vacuum polarization 
pairs act as mediators of an 
indirect interaction between the field of 
the antineutron magnetic moment and the electric field of the nucleus.  
A necessary consequence is that the 
antineutron is able to test some 
features of the nuclear electric field. Saying this in other words, 
some electric multipoles are induced in the antineutron. 


However unusual it may sound, the prediction that the magnetic field and 
the electric field produced by different particles  
may be coupled by vacuum 
polarization just belongs to the class of  
photon-photon interactions, like $\gamma\gamma$ $\rightarrow$ 
$X$ experiments at $e^+e^-$ colliders, for which decades of 
experimental results exist since \cite{Adone,Pluto,NSB}. 
The important difference is that in the collider case the local 
fields are so intense to permit vacuum fluctuations into 
real $\bar{q}q$ pairs with involvement of the strong interactions and 
large enhancement of the cross sections. 
So a sensitivity of the antineutron to the nuclear electric 
field cannot be excluded, the real problem being its strength. 

The induced dipole value given by eq.29 of \cite{Dominguez09} 
depends on a length parameter ``$a$'' (the dipole is $\propto$ $1/a^3$). 
In \cite{Dominguez09} $a$ parameterized 
a charge current distribution in the neutron, and values of $a$ 
$\leq$ 1 fm were suggested. In a later work \cite{Zimmer12} $a$ 
was reinterpreted as a parameter 
incorporating the effects of higher order corrections  
and values of $a$ over 7 fm were suggested. 
Given the value of the electric field of the Sn 
nucleus, we obtain the dipole value fitted by us for 
$a$ $\approx$ 2 fm. 

\subsection{Nonlinearity} 
\label{nonlinearity}

As fitted by us, the difference between the polarizabilities in the Cu 
and Sn cases means that we are out of a linear regime, where the 
induced dipole would be proportional to the inducing field. 
In our case  a stronger field induces a dipole that is weaker than 
a linear law expectation. We may speak of ``strong-field saturation''. 

If the induced dipole 
is associated to a 
stretching of the projectile internal 
structure this saturation is natural, since QCD 
interactions between the hadron constituents are well known to be 
small at distances $<<$ 1 fm but rapidly increasing at distances 
$\sim$ 1 fm. These would oppose to an antineutron dipole with 
strength $e\delta$ for $\delta$ $>$ 1 fm, saturating the polarization at 
large $\delta$. More in general, it is unavoidable that a force 
acting on the antineutron as a whole brings some stress on 
the internal structure of this hadron. When this stress implies 
deformations of size $>$ 1 fm, QCD interactions will introduce 
nonlinearity.   

Within QED, the effects 
as calculated by \cite{Dominguez09} correspond to a given power of 
$\alpha$ (see eq.2 in that work). 
Extending to higher orders would introduce some degree of negative 
feedback at increasing fields because of Lenz's law. This would 
naturally show as a smaller polarizability in presence of a larger 
inducing field. This nonlinearity must be expected for $E$ $>>$ 
10$^{18}$ V/m (the Sauter's critical value) because then the $e^+e^-$ 
pairs proliferate instead of being a small perturbation, and their 
reciprocal screening cannot be neglected. 

In conclusion, it is reasonable to expect nonlinearity when the 
dipole value corresponds to $eL$ with $L$ $>$ 1 fm because of QCD 
confinement. It is also reasonable to expect nonlinearity 
when the inducing field is $>>$ 
10$^{18}$ V/m because of QED higher order terms and Lenz's law. 
And in both cases this nonlinearity is expected to lead towards 
saturation. 

\section{Conclusions and perspectives}
\label{conclusion}

The main conclusion of this note is that a medium range force, 
in our assumption of electromagnetic origin, allows for a physically 
based justification of the $1/p^2$ term in the fit of the antineutron 
annihilation cross section on medium and large nuclei 
at 76-375 MeV/c. Exploiting recent analysis claiming 
the existence of an induced electric dipole moment in the neutron, we 
have assumed that an analogous moment is induced in an antineutron by 
the electric field of 
a nearby nucleus. The attracting force deriving from this leads to a steep  
energy dependence of the annihilation cross section for projectile 
momenta  below 100 MeV/c. We have presented arguments to 
demonstrate that a medium range interaction is necessary for justifying 
the steep rise of the cross section in the region 50-200 MeV/c. 

In perspective, our work suggests the need for 1-2 more antineutron 
data points that permit to estimate  
the exact slope of the $\bar{n}$-nucleus cross section in the 
region 50-100 MeV/c, and understand where exactly an $a + b/p$ law 
must be abandoned. This would put more selective constraints on 
the form (strength, range and asymptotic behavior) of the involved 
medium-range interaction term. 

An analysis of elastic data, as suggested in \cite{Zimmer12}, 
could also shed light on these points. In the energy region considered 
in this work, data on the differential distribution of elastically scattered 
antinucleons are still absent. 

A relevant open point with these data, presently admitting 
neither an intuitive justification nor one from a complex model, is the 
relative size of the $\bar{n}$-nucleus and $\bar{p}$-nucleus 
annihilation cross sections for $p$ $<$ 
100 MeV/c. Since in this moment the antineutron data in this region 
are more systematic and precise than the antiproton ones, 
an as detailed as possible exploration of 
the antiproton annihilation on nuclei below 100 MeV/c is required. 
In particular what is missing is a detailed interpolation between the 
behavior of the antiproton annihilation on $light$ nuclei like hydrogen 
or helium (where a very steep Coulomb rise is visible, and antiproton cross 
sections overcome antineutron ones by a factor as large as three) and 
on medium and large nuclei where we find cross sections that seem 
far smaller than expected. Ideally, a precision that allows to 
distinguish constant, $1/p$ and $1/p^2$ components of the cross section 
$p$-dependence is required.




\begin{thebibliography}{99}

\bibitem{Baker06} 
C.A.Baker et al, 
Phys.Rev.Lett.{\bf 97}, 131801 (2006). 

\bibitem{Fedorov10}
V.V.Fedorov et al, 
Phys.Lett.{\bf B 694}, 22-25 (2010) 

\bibitem{Dar00}
S.Dar, arXiv:hep-ph/0008248. 

\bibitem{Pospelov05}
M.Pospelov and A.Ritz, 
Annals Phys.318:119-169,2005

\bibitem{Peng08}
J.C.Peng, 
Mod.Phys.Lett. {\bf A23}, 1397-1408 (2008).  

\bibitem{Lamoreaux09} 
S.K. Lamoreaux and R. Golub, J. Phys. {\bf G
36} 104002 (2009). 


\bibitem{Dominguez09} 
C.A.Dominguez, H.Falomir, M.Ipinza, S.Kohler, M.Loewe, and J.C.Rojas, 
Phys.Rev.{\bf D 80}, 033008 (2009).  

\bibitem{Sauter31}
F. Sauter, Z. Phys. {\bf 69}, 742-764 (1931). 
\bibitem{Zimmer12} 

O.Zimmer, C.A.Dominguez, H.Falomir, and M.Loewe, 
Phys.Rev.{\bf D 85}, 013004 (2012).  

\bibitem{Gunderson81}
B. Gunderson et al., Phys. Rev. D 23 (1981) 587.

\bibitem{Agnello88}
M. Agnello et al., Europhys. Lett. 7 (1988) 13.

\bibitem{Barbina97}
C. Barbina et al., Nucl. Phys. A 612 (1997) 346.

\bibitem{Ableev94}
V.G. Ableev et al., Nuovo Cimento 107 A (1994) 943.


\bibitem{Astrua02}
M. Astrua et al., Nucl. Phys. A 697 (2002) 209–224

\bibitem{Bressani03}
T.Bressani and A.Filippi, Phys. Rep. {\bf 383} 213–297 (203). 



\bibitem{bizzarri1974}
 R. Bizzarri et al., Nuovo Cim. A22 (1974) 225

\bibitem{kalogeropoulos1980}
T.E. Kalogeropoulos and G.S. Tzanakos, Phys. Rev. 22D (1980) 2585

\bibitem{balestra1984}
F. Balestra et al., Phys. Lett. B 149 (1984) 69

\bibitem{balestra1985}
F. Balestra et al., Phys. Lett. B 165 (1985) 265

\bibitem{balestra1986}
F. Balestra et al., Nucl. Phys. A 452 (1986) 573

\bibitem{balestra1989}
F. Balestra et al., Phys. Lett. B 230 (1989) 36

\bibitem{bruckner1990}
W. Br\"uckner et al., Z. Phys. A 335 (1990) 217

\bibitem{bertin1996}
A. Bertin et al. (OBELIX Collaboration), Phys. Lett. B 369 (1996) 77

\bibitem{benedettini1997}
A. Benedettini et al. (OBELIX Collaboration), Nucl. Phys. B Proc. Suppl. 56A (1997) 58

\bibitem{zenoni1999_1} A. Zenoni et al. (OBELIX Collaboration), Phys. Lett. B 461 (1999) 405

\bibitem{zenoni1999_2} A. Zenoni et al. (OBELIX Collaboration), Phys. Lett. B 461 (1999) 413

\bibitem{bianconi2000_1} A. Bianconi et al., Phys. Lett. B 481 (2000) 194

\bibitem{bianconi2000_2} A. Bianconi et al., Phys. Lett. B 492 (2000) 254

\bibitem{Bianconi11} 
A. Bianconi et al., Phys. Lett. B 704 (2011) 461.


\bibitem{LL3}
L. D. Landau, E. M. Lifshitz, ``Course of Theoretical Physics -
Quantum Mechanics (Non-Relativistic Theory)''
Butterworth-Heinemann Vol 3

\bibitem{Bianconi00}
A.Bianconi, G.Bonomi, E.Lodi Rizzini, L.Venturelli and A.Zenoni, 
Phys.Rev. {\bf C62} 014611 (2000). 

\bibitem{Bianconi00b}
A.Bianconi, G.Bonomi, M.P.Bussa, E.Lodi Rizzini, L.Venturelli and A.Zenoni, 
Phys.Lett. {\bf B 483} 353-359 (2000).  

\bibitem{Batty01}
C.J.Batty, E.Friedman, and A.Gal, 
Nucl.Phys. {\bf A 689} 721-740 (2001). 


\bibitem{Carbonell93}
J.Carbonell and K.V.Protasov, Hyp. Int. {\bf 76} 327 (1993). 

\bibitem{Friedman14}
E.Friedman, Nucl. Phys. {\bf A 925} (2014) 141–149

\bibitem{Iazzi00}
F.Iazzi, et al., Phys. Lett. B 475 (2000) 378.

\bibitem{Adone}
C.Bacci et al, Proc. First EPS conf. on meson resonances and related 
electromagnetic phenaoena, Bologna, 1971. 
C.Bacci et al, Lett. Nuovo Cim. 3, 709 (1972). 

\bibitem{Pluto}
C.H.Berger et al, Phys.Lett. B, 120 (1979). 

\bibitem{NSB}
S.E.Baru et al, Z.Phys.C 53 219 (1992). 

















\end{thebibliography}
\end{document}